\documentclass[journal]{IEEEtran}
\usepackage{cite}

\usepackage{multirow}

\ifCLASSINFOpdf
  \usepackage[pdftex]{graphicx}
  \graphicspath{{fig/pdf/}{fig/jpeg/}{fig/png/}}
  \DeclareGraphicsExtensions{.pdf,.jpeg,.png}
\else
  \usepackage[dvips]{graphicx}
  \graphicspath{{fig/eps/}}
  \DeclareGraphicsExtensions{.eps}
\fi
\usepackage{tikz}

\newcommand\submittedtext{%
  \footnotesize This work has been submitted to the IEEE for possible publication. Copyright may be transferred without notice, after which this version may no longer be accessible.}

\newcommand\submittednotice{%
\begin{tikzpicture}[remember picture,overlay]
\node[anchor=south,yshift=10pt] at (current page.south) {\fbox{\parbox{\dimexpr0.65\textwidth-\fboxsep-\fboxrule\relax}{\submittedtext}}};
\end{tikzpicture}%
}

\usepackage{amsmath}
\usepackage{amssymb} 

\usepackage{array}
\DeclareUnicodeCharacter{03BC}{}

\begin{document}
%
\title{A Learning Method \\ for Optimal Chirp Predistortion}
%
%
%

\author{Averal N. Kandala
        and~Ali M. Niknejad
\thanks{This work was supported by the National Science Foundation (NSF) ACED Fab and Graduate Research Fellowship programs under award numbers ECCS 2314969 and DGE 2146752, respectively. Additional support was provided by the Center for Ubiquitous Connectivity (CUbiC), sponsored by the Semiconductor Research Corporation (SRC) and Defense Advanced Research Projects Agency (DARPA) under the JUMP 2.0 program.}
\thanks{Averal N. Kandala and Ali M. Niknejad are with the Department
of Electrical Engineering and Computer Sciences, University of California, Berkeley, Berkeley, 
CA 94720, USA. Correspondence: averal@berkeley.edu.}

}

%
%

\markboth{Preprint}%
{Kandala and Niknejad: A Learning Method for Optimal Chirp Predistortion}
%



\maketitle


\submittednotice

\begin{abstract}
This article introduces the ``chart \& chirp" method of one-shot, in situ VCO tuning curve estimation, learning, and predistortion, which provides an alternative to prevailing LMS-based background calibration loops. The proposed approach utilizes a cycle-counting FDC to estimate the VCO tuning curve and a linearly interpolating QDAC to actuate the predistortion. An LLSE-based learning process generates predistorted DAC control codes for any physically achievable chirp parameters. In simulation, 50 and 51 kHz of RMS FM error is achieved for 5 and 20 \textmu s chirps, respectively. The locations of nonlinearity-induced spurs in the IF spectrum are predicted via analysis of the DFT of FM error in generated chirps, with higher frequency DAC updates weakening these spurs and moving them out of the IF band of interest. Simulation in a coherent, monostatic radar model reveals that deterministic phase error from the chirp nonlinearity arises at a level above -70 dBc/Hz at 1 MHz offset, dominating random phase noise and setting the IF SNDR for $R \leq$ 23 m and $R = $ 92 m around 40 dB and 24 dB, respectively.
\end{abstract}


%
\IEEEpeerreviewmaketitle

\section{Introduction}
\begin{figure}[!t]
\centering
\includegraphics[width=3.5in]{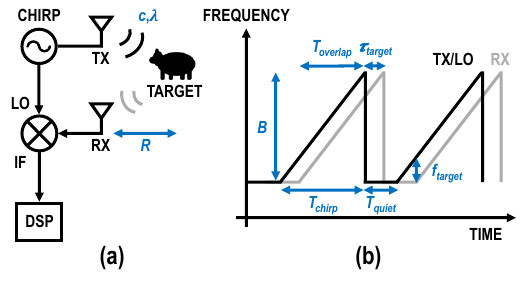}
\caption{(a) Simplified monostatic FMCW radar conceptual diagram, as modeled in this work. Amplifiers, filters, and frequency multipliers are typically added for signal conditioning and translation. An ADC precedes the DSP. (b) FMCW sawtooth chirp waveforms, with relevant annotations.}
\label{fig1}
\end{figure}

%
%
%
%
\IEEEPARstart{T}{he} sensing performance of a monostatic frequency-modulated continuous wave (FMCW) radar system is determined by the linear FM ``chirp" used at both its transmit (TX) and coherent local oscillator (LO) ports (Fig. \ref{fig1}). A sawtooth chirp $f_{chirp}(t)$ beginning at frequency $f_{0}$ can be described over time as follows:

\begin{equation}
    f_{chirp}(t) = f_{0} + \frac{B}{T_{chirp}} \cdot t\label{eq:chirp_def}
\end{equation}

The chirp is defined by its modulation bandwidth, $B$, and duration, $T_{chirp}$. The ratio of $B$ to $T_{chirp}$, referred to as the ``chirp slope", prescribes the distance $R$ to the target via a frequency, $f_{target}$, measured at the intermediate frequency (IF), or baseband, port:

\begin{equation}
    f_{target} = \frac{B}{T_{chirp}} \cdot \tau_{target} = \frac{B}{T_{chirp}} \cdot \frac{2R}{c} \label{eq:target_freq}
\end{equation}

In \eqref{eq:target_freq}, $\tau_{target}$ is the round-trip time delay to the target and back. The chirp parameters also establish the system's minimum range resolution, $\Delta R_{min}$, and maximum unambiguous observable velocity, $v_{max}$, respectively \cite{rao_introduction_nodate}:


\begin{equation}
    \Delta R_{min} = \lim_{T_{obs} \to T_{chirp}} \frac{1}{T_{obs}} \cdot \frac{c}{2} \cdot \frac{T_{chirp}}{B} = \frac{c}{2B}\label{eq:range_res}
\end{equation}

\begin{equation}
    |v_{max}| = \frac{\lambda}{4(T_{chirp} + T_{quiet})}\label{eq:max_vel}
\end{equation}

\begin{figure*}[!t]
\centering
\includegraphics[width=7.16in]{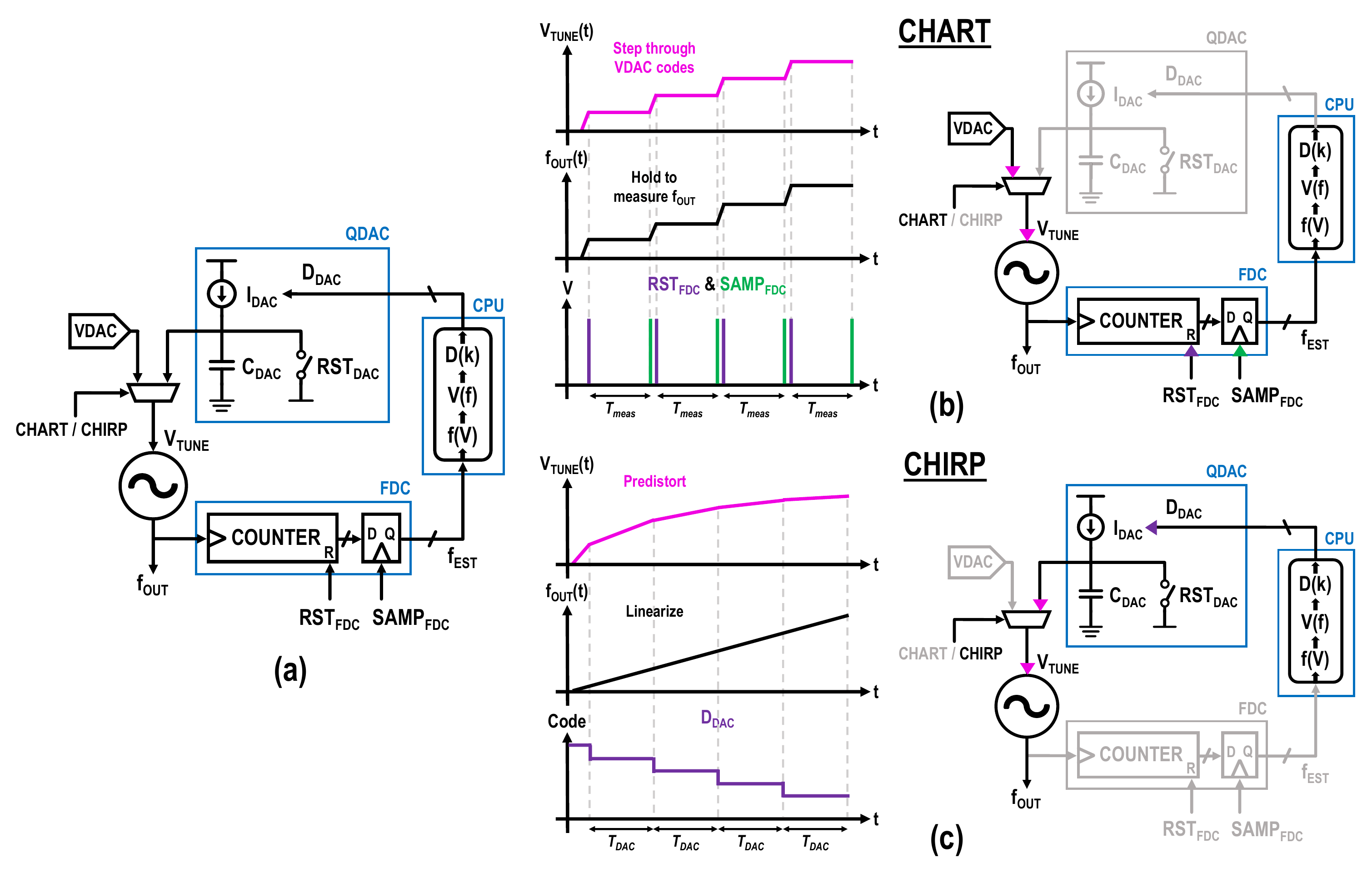}
\caption{(a) Conceptual diagram of the proposed system. System operation and relevant waveforms during (b) ``chart" and (c) ``chirp" phases.}
\label{fig2}
\end{figure*}
 
Here, $T_{obs}$ is the IF observation interval, maximized at $T_{overlap}$; $c$ is the speed of light; $\lambda$ is the wavelength; and $T_{quiet}$ is the interval between chirps (Fig. \ref{fig1}). To optimize performance in light of \eqref{eq:range_res} and \eqref{eq:max_vel}, modern FMCW radars seek to maximize $B$ (``wide tuning") and minimize $T_{chirp}$ (``fast chirping"). This consequently raises the IF bandwidth through \eqref{eq:target_freq}, mitigating receiver (RX) flicker noise obscuration of nearby targets at the potential expense of faster required sampling in the analog-to-digital converter (ADC) \cite{deng_self-adapted_2022-1}. Meanwhile, the signal-to-noise-and-distortion ratio (SNDR) at the IF port is directly degraded by chirp nonlinearity and phase noise (PN) \cite{ayhan_impact_2016-1, siddiq_phase_2019-1}.

Due to their low phase noise, LC voltage-controlled oscillators (VCOs) are preferred for chirp synthesis. Frequency tuning occurs through the capacitance, either in the discrete steps of a switched-capacitor bank (``cap bank"), or in the modulation of analog varactors, or both. Varactors provide more linear continuous tuning at the expense of tuning range. In either case, chirp nonlinearity results from a combination of control-to-capacitance and capacitance-to-frequency nonlinearity across the tuning range \cite{renukaswamy_12-mw_2020-2}. In increasing order of precision, this is addressed either through feedback control, such as in a phase- or frequency-locked loop (PLL, FLL), through auxiliary varactor compensation, or through calibrated predistortion of the VCO control. 

Although closed-loop (``low pass") methods are traditionally preferred for phase noise mitigation, they begin to fail as the chirp duration decreases to the order of the loop settling time and below. As a result, direct (``high pass") modulation of the VCO is typically employed for $T_{chirp} < 50 \, \text{\textmu s}$ \cite{deng_self-adapted_2022-1, zhang_276_2026, tesolin_10-ghz_2024-2, renukaswamy_12-mw_2020-2, renukaswamy_16-ghz_2024-2, cherniak_23-ghz_2018-1, shi_self-calibrated_2019-4, shen_24_2022-1, wang_11-ghz_2026}, with or without an additional loop to lock the phase to a reference, requiring nonlinearity compensation or control predistortion. 

Nonlinearity compensation can be achieved for continuous VCO tuning using a varactor array formed of elements exhibiting differing and complementary tuning characteristics. This approach has been demonstrated in the use of multiple varactors at different bias levels \cite{zhao_high-linearity_2025}, as well as through the combination of NMOS and PMOS varactors with inherently complementary nonlinearities \cite{zhang_276_2026}. As this form of compensation relies on calibration of arbitrary characteristics without mathematically exact nonlinearity cancellation, explicit predistortion of the VCO control is preferred to further minimize chirp nonlinearity. To implement this, a background calibration loop can update look-up table (LUT) values for the frequency control over multiple chirps, typically via the least mean square (LMS) algorithm \cite{shi_self-calibrated_2019-4, renukaswamy_12-mw_2020-2, renukaswamy_16-ghz_2024-2, cherniak_23-ghz_2018-1, zhang_276_2026}. This approach can be augmented by leveraging prior knowledge of the VCO tuning curve (analytical and simulated) to improve accuracy \cite{tesolin_10-ghz_2024-2}. For these implementations, the settings required for algorithm convergence are derived heuristically or empirically rather than analytically, and recalibration (or more LUT memory) is required to change the modulation parameters.

To address these key challenges, this work proposes a two-phase ``chart \& chirp" method of one-shot, in situ VCO tuning curve estimation and learning (``chart") with on-demand predistortion (``chirp") for all physically realizable chirp parameters. Section II presents the proposed system architecture and illustrates the ``chart \& chirp" estimation, learning, and predistortion procedure. Section III analyzes the spectra of the chirp FM error and IF output based on chirp nonlinearity and phase noise theory \cite{ayhan_impact_2016-1, siddiq_phase_2019-1}. Section IV includes a discussion of simulation results for a coherent, monostatic radar model. Section V summarizes the conclusions of the work.






\section{System Concept}

The proposed system is illustrated in Fig. \ref{fig2}. It consists of four essential components: 1) a continuous-tuning (analog) VCO; 2) an FDC for frequency estimation; 3) a central processing unit (CPU), with associated LUT, for learning and predistortion; and 4) an integrating DAC (QDAC) for low-noise, linear interpolation of the applied predistortion. As detailed in Figs. \ref{fig2}(b) and \ref{fig2}(c), an analog multiplexer selects the VCO control voltage based on the phase of operation. In the ``chart" phase, a voltage DAC (VDAC) linearly steps through the VCO's tuning range, while the FDC estimates the VCO frequency at each step. This establishes a voltage-to-frequency mapping, which the CPU then converts into a sequence of predistorted control codes based on the QDAC and desired modulation parameters. In the ``chirp" phase, the QDAC applies the predistortion to the VCO control voltage to produce a linear chirp. An additional loop to lock the phase is not depicted but can potentially be added.

\subsection{Frequency Estimation}

\begin{figure}[!t]
\centering
\includegraphics[width=3.5in]{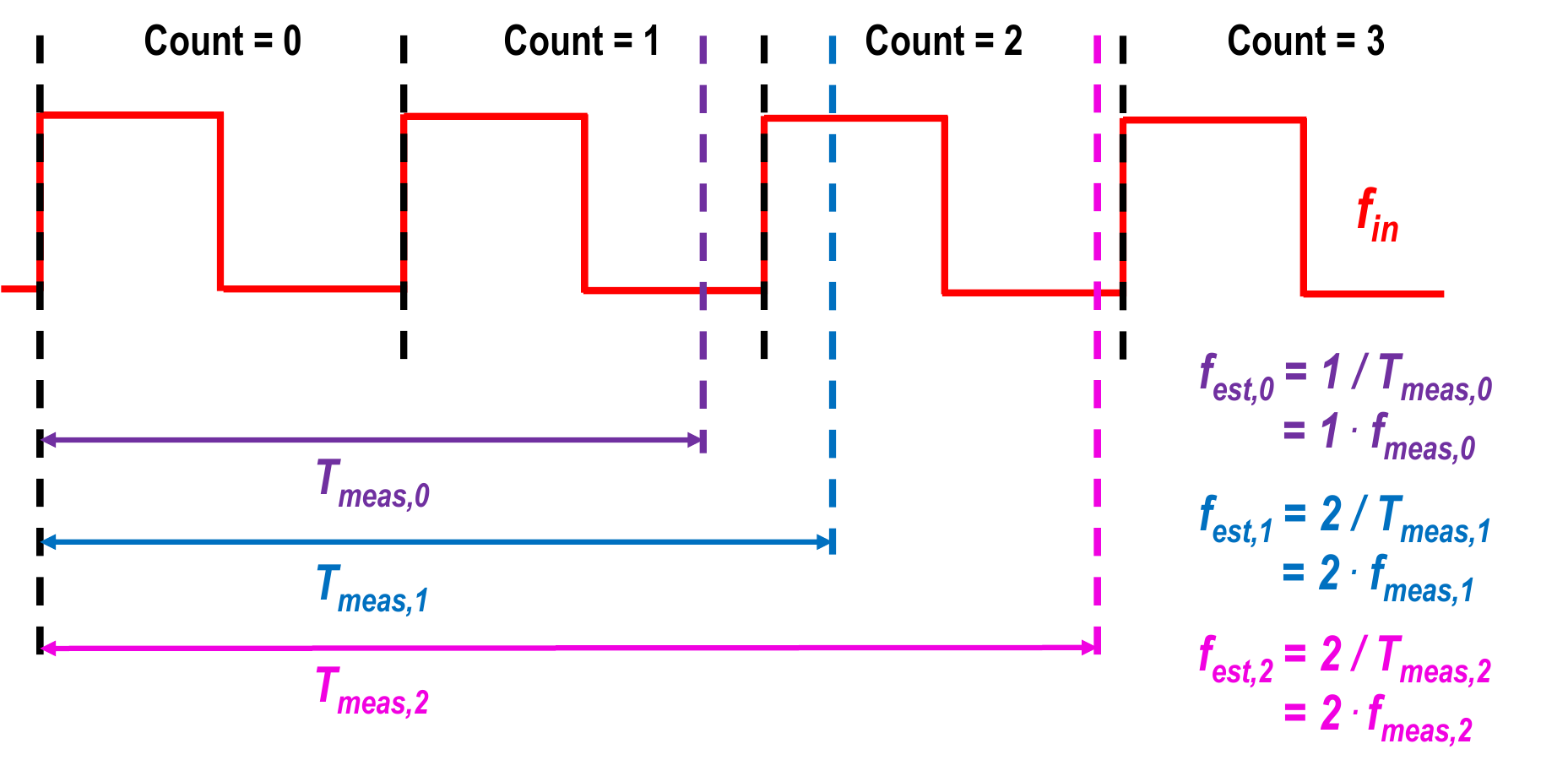}
\caption{Frequency estimation through timed measurement of the counter output. FDC resolution improves as $T_{meas}$ increases. Ideally, $f_{meas} \ll f_{in}$.}
\label{fig3}
\end{figure}

The conceptual operation of the FDC is illustrated in Fig. \ref{fig3}. To estimate the VCO frequency, the rising edges of the VCO output are counted over a measurement interval, $T_{meas}$ \cite{sadhu_linearized_2013}:

\begin{equation}
    f_{est} = \frac{N_{count}}{T_{meas}} = N_{count} \cdot f_{meas}\label{eq:f_est}
\end{equation}

The estimated frequency is $f_{est}$, the count is $N_{count}$, and $f_{meas} = 1/T_{meas}$. In \eqref{eq:f_est}, the resolution of the frequency estimation is set to $f_{meas}$ by the duration of the measurement, rather than by any intrinsic property of the counter circuit. If the counter overflows in $T_{overflow}$, any resolution down to $1 / T_{overflow}$ can be supported. Put another way, for a given input frequency, $f_{in}$, the counter must support the following condition:

\begin{equation}
    N_{count,max} = \lfloor f_{in} \cdot T_{overflow} \rfloor > \frac{f_{in}}{f_{meas}}\label{eq:count_max}
\end{equation}

Expressing the capacity of the counter in bits as $N_{bits,FDC}$, the minimum achievable resolution of the FDC can be derived:

\begin{equation}
    N_{count,max} = 2^{N_{bits,FDC}} - 1\label{eq:count_bits}
\end{equation}

\begin{equation}
    f_{meas,min} = \frac{f_{in}}{N_{count,max}} =  \frac{f_{in}}{2^{N_{bits,FDC}} - 1}\label{eq:count_min_res}
\end{equation}

In the limit that $N_{bits,FDC}$ goes to infinity, the minimum FDC resolution theoretically goes to zero. As this method of frequency estimation is an averaging process, this result is in line with the following consequence of the law of large numbers: $f_{est}$ approaches $f_{in}$ in the limit that $T_{meas} \gg 1/f_{in}$ (or $f_{meas} \ll f_{in}$). 

Therefore, to achieve fine resolution in the FDC, $T_{meas}$ should be made as high as the capacity of the counter will allow. The FDC measurement procedure is shown in Fig. \ref{fig2}(b). The measurement interval begins when the counter is reset and ends when the counter output is sampled by a bank of D flip-flops (DFFs or registers). To improve measurement accuracy against sampling errors, multiple measurements can be taken for the same input frequency. Outlier measurements can be identified and discarded, and the remaining values can be averaged. 


\begin{figure}[!t]
\centering
\includegraphics[width=3.5in]{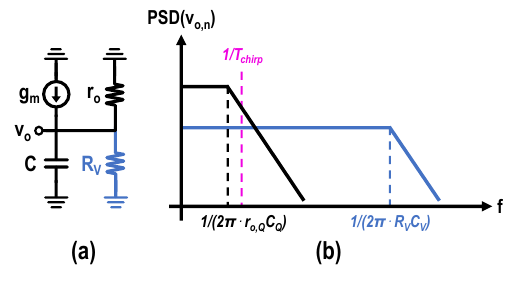}
\caption{(a) Small signal model for the QDAC (black) and VDAC (black and blue). (b) Output thermal voltage noise power spectral density (PSD) over frequency for the QDAC (black) and VDAC (blue).}
\label{fig6}
\end{figure}

\subsection{QDAC Properties}

Fig. \ref{fig2} presents a simplified version of the QDAC concept \cite{sakurai_15ghz-modulation-range_2011, renukaswamy_12-mw_2020-2}, suitable for producing ``up" or sawtooth chirps. The current DAC (IDAC) output can be mirrored and switches can be added to the output to enable a ``down" chirp phase, allowing triangular chirp modulation as well. The QDAC topology provides two key benefits compared to a voltage DAC (VDAC): low noise (and power) and linear interpolation. 

\subsubsection{Noise}

The AC small-signal models for the two topologies are almost identical (Fig. \ref{fig6}(a)) \cite{renukaswamy_12-mw_2020-2}. In both cases, a current source with transconductance, $g_m$, and output resistance, $r_o$, drives a capacitance, $C$ (with subscripts $Q$ and $V$ for the QDAC and VDAC, respectively). In the VDAC, the large-signal output voltage is defined as the IR drop across a resistor, $R_V$, in shunt with $C_V$ \cite{shi_self-calibrated_2019-4}. As a result, $C_V$ is intended to be small to minimize the VDAC settling time. In contrast, the QDAC output is set by linear slewing across $C_Q$, typically allowing $C_Q \gg C_V$ \cite{renukaswamy_12-mw_2020-2}. A comparison can be established by analyzing the thermal voltage noise at the output in CMOS \cite{dastgheib_calculation_2008}:

\begin{equation}
    \overline{v_{o,n,Q}^2} = \int_{0}^{\infty} 4kT \gamma g_{m,Q}\left \lvert \frac{r_{o,Q}}{1 + j \omega r_{o,Q} C_Q}\right \rvert^2  \,df \label{eq:calc_noise_Q}
\end{equation}

\begin{equation}
    \overline{v_{o,n,Q}^2} =  \frac{kT}{C_Q}\cdot (\gamma \cdot g_{m,Q} r_{o,Q})\label{eq:total_noise_Q}
\end{equation}

\begin{equation}
    \overline{v_{o,n,V}^2} \approx \int_{0}^{\infty} 4kT ( \gamma g_{m,V} + \frac{1}{R_V} ) \left \lvert \frac{R_V}{1 + j \omega R_V  C_V}\right \rvert^2  \,df \label{eq:calc_noise_V}
\end{equation}

\begin{equation}
    \overline{v_{o,n,V}^2} \approx \frac{kT}{C_V}\cdot (\gamma \cdot g_{m,V} R_V + 1)\label{eq:total_noise_V}
\end{equation}

Here, $f$ is the frequency, $\omega = 2\pi f$, $k$ is the Boltzmann constant, $T$ is the absolute temperature, $\gamma$ is a CMOS device parameter, and it is assumed that $R_V \ll r_{o,V}$, such that $r_{o,V}||R_V \approx R_V$. For similar operating conditions, when $\frac{C_Q}{C_V} > \frac{r_{o,Q}}{R_V}$, the QDAC is lower noise. It is possible to choose $R_V$ small and $C_V$ large such that the VDAC has comparable or even superior noise. However, this comes at a significant power penalty, with a factor of 20 difference reported between \cite{renukaswamy_12-mw_2020-2} (0.5 mW) and \cite{shi_self-calibrated_2019-4} (10 mW). Indeed, the lack of a DC current path to ground in the QDAC output makes it naturally lower power than the VDAC.

In \eqref{eq:calc_noise_Q} and \eqref{eq:calc_noise_V}, the noise integration bandwidth is $[0, \infty)$. During chirp modulation, the relevant noise integration bandwidth will be $[1/T_{chirp}, \infty)$ (Fig. \ref{fig6}(b)). As a result, the high-noise regime of the QDAC at frequencies below $1/(2\pi\cdot r_{o,Q}C_Q)$ can be avoided entirely, especially for fast chirps. The VDAC sees less benefit since $R_V$ and $C_V$ are both exclusively smaller than their QDAC counterparts ($r_{o,Q}$ and $C_Q$), resulting in a much higher filter corner frequency at around $1/(2\pi\cdot R_{V}C_V)$. Less flicker noise is observed in either case for low $T_{chirp}$.

\subsubsection{Linear Interpolation}

The more fundamental and significant advantage of the QDAC lies in the fact that it naturally performs linear interpolation at its output (Fig. \ref{fig2}(c)). 

\begin{equation}
    \frac{d V_{tune}(t)}{dt} = \frac{I_{DAC}(t)}{C_{DAC}} = \frac{I_{LSB}}{C_{DAC}} \cdot D_{DAC}(t) \label{eq:dac_slope}
\end{equation}

\begin{equation}
    V_{tune}(t) = V_{tune}(0) + \frac{I_{LSB}}{C_{DAC}} \cdot \int_{0}^{t} D_{DAC}(\tau) \,d \tau \label{eq:dac_voltage}
\end{equation}

Equation \eqref{eq:dac_slope} sets the slope of the VCO tuning voltage over time, $V_{tune}(t)$, to the ratio of the IDAC current, $I_{DAC}(t)$, and the QDAC output capacitance, $C_{DAC}$. $I_{DAC}(t)$ can be factored as the product of $I_{LSB}$, the IDAC resolution (LSB), and $D_{DAC}(t)$, the IDAC control code over time. As will be expanded in the next section, \eqref{eq:dac_voltage} enables perfect linear DAC interpolation of any set of desired tuning voltage points based on a given starting voltage and a sequence of IDAC control codes (assuming infinite IDAC precision and full-scale range). This significantly improves chirp linearity compared to the zero-order hold (ZOH) DAC interpolation resulting from the discrete VCO frequency steps of a digitally-controlled cap bank or a varactor tuned by a VDAC. 

\subsection{Learning, Predistortion, and Linearization}

\begin{figure*}[!t]
\centering
\includegraphics[width=7.16in]{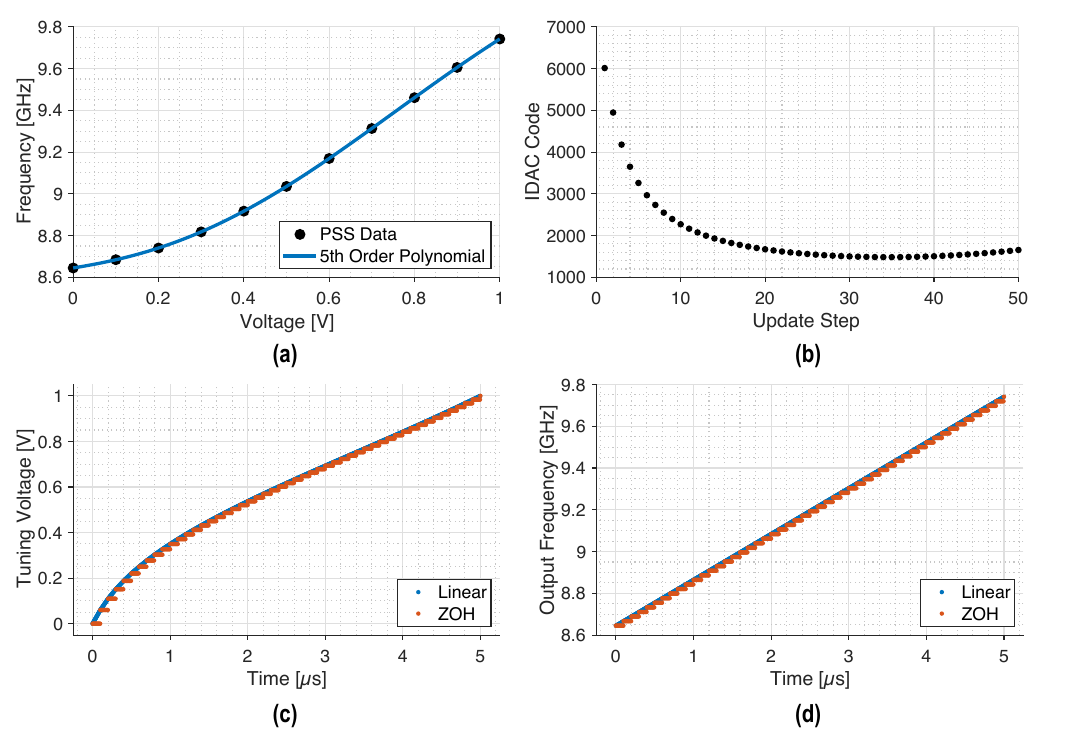}
\caption{(a) Simulated VCO tuning curve, (b) predistorted IDAC codes, (c) predistorted tuning voltage with linear and ZOH DAC interpolation, and (d) predistorted chirp with linear and ZOH DAC interpolation.}
\label{fig7}
\end{figure*}

As illustrated in Fig. \ref{fig2}(b), a voltage-to-frequency (``forward") mapping, $\hat{f}(V)$, is established in the ``chart" phase as a VDAC steps through the tuning range of the VCO and the FDC provides an estimate of the frequency at each voltage step. Since the settling time of this VDAC is not critical, a simple topology, such as the R-2R ladder with a large filtering capacitance at its output \cite{marche_modeling_2010}, can be used. 

\subsubsection{Learning}

Taking the frequency estimate, $\hat{f}$ as the independent variable and the voltage, $V$, as the dependent variable, a frequency-to-voltage (``backward") mapping can be constructed using the linear least-squares estimation (LLSE) or linear regression method, specifically ordinary least-squares (OLS):

\begin{equation}
    \mathbf{V} = \mathbf{\hat{F}} \mathbf{b} \label{eq:llse_basic}
\end{equation}

\begin{equation}
    \mathbf{V} =
    \begin{bmatrix}
        V[0] &
        V[1] &
        \ldots &
        V[N]
    \end{bmatrix}^T
    \label{eq:llse_V}
\end{equation}

\begin{equation}
    \mathbf{\hat{F}} = \begin{bmatrix} \mathbf{\hat{F}_{P}} & \mathbf{\hat{F}_{1/P}} \end{bmatrix} \label{eq:llse_f_hat}
\end{equation}

\begin{equation}
    \mathbf{\hat{F}_{P}} = 
    \begin{bmatrix}
        1 & \hat{f}[0] & \hat{f}[0]^2 & \ldots & \hat{f}[0]^P \\
        1 & \hat{f}[1] & \hat{f}[1]^2 & \ldots & \hat{f}[1]^P \\
        \vdots & \vdots & \vdots & \ddots & \vdots \\
        1 & \hat{f}[N] & \hat{f}[N]^2 & \ldots & \hat{f}[N]^P
    \end{bmatrix}
    \label{eq:llse_f_hat_p}
\end{equation}

\begin{equation}
    \mathbf{\hat{F}_{1/P}} = 
    \begin{bmatrix}
        \hat{f}[0]^{1/2} & \hat{f}[0]^{1/3} & \ldots & \hat{f}[0]^{1/P} \\
        \hat{f}[1]^{1/2} & \hat{f}[1]^{1/3} & \ldots & \hat{f}[1]^{1/P} \\
        \vdots & \vdots & \ddots & \vdots \\
        \hat{f}[N]^{1/2} & \hat{f}[N]^{1/3} & \ldots & \hat{f}[N]^{1/P}
    \end{bmatrix}
    \label{eq:llse_f_hat_1_p}
\end{equation}

\begin{equation}
    \mathbf{b}=
    \begin{bmatrix}
        b_{0} &
        b_{1} &
        b_{2} &
        \ldots &
        b_{P} &
        b_{1/2} &
        \ldots &
        b_{1/P}
    \end{bmatrix}^T
    \label{eq:llse_b}
\end{equation}

\begin{equation}
    \mathbf{\hat{b}} = (\mathbf{\hat{F}}^T\mathbf{\hat{F}})^{-1}(\mathbf{\hat{F}}^T\mathbf{V}) \overset{\mathrm{def}}{=} \mathbf{V} \backslash \mathbf{\hat{F}}
    \label{eq:llse_b_hat}
\end{equation}

Here, $N = N_{chart}$, representing the integer number of voltage increments, and $P$ (typically $\geq 3$) is the (selected) positive integer order of the fit. Therefore, $N_{chart} + 1$ is the number of charted points and the length of the $\hat{f}$ vector. The $(N_{chart} + 1) \times (2P)$ design (or feature) matrix, $\mathbf{\hat{F}}$, represents the following assumed backward model for the VCO:

\begin{equation}
    \begin{aligned}
    V(f) = b_0 + b_1f + b_2f^2+ \ldots +b_Pf^P \\ 
    + b_{1/2}f^{1/2} + \dots +b_{1/P}f^{1/P}
    \end{aligned}
    \label{eq:llse_backward_model}
\end{equation}

As a result, $\mathbf{\hat{b}}$ is the OLS estimate of the backward model parameter vector, $\mathbf{b}$. This model contains regular polynomial terms up to order $P$ (in $\mathbf{\hat{F}_{P}}$), as well as fractional polynomial terms with exponents down to $1/P$ (in $\mathbf{\hat{F}_{1/P}}$). These fractional polynomial terms are chosen to reciprocate a simple order-$P$ polynomial model for the VCO in the forward direction:

\begin{equation}
    \begin{aligned}
    f(V) = a_0 + a_1V + a_2V^2+ \ldots +a_PV^P
    \end{aligned}
    \label{eq:llse_forward_model}
\end{equation}

To ensure a proper fit to the polynomial model, a starting condition of $(V[0], \hat{f}[0]) = (0, 0)$ is enforced by subtracting the true $V[0]$ and $\hat{f}[0]$ from the relevant input vectors. For numerical stability, large values are avoided in $\hat{f}$ by expressing the values in GHz, rather than Hz. This scaling can be adjusted based on the VCO and values in question. 

\subsubsection{Predistortion}

The backward model estimate, $\mathbf{\hat{b}}$, can now be used to generate the optimal predistorted voltage curve, $\mathbf{V_{PD}}$, from an ideal, perfectly linear chirp in frequency, $f_{lin}$, adjusted appropriately to start at zero (as discussed above): 

\begin{equation}
    f[k] = f_{lin}[k] = \frac{B_{des}}{N_{DAC}} \cdot k \;\;\; \forall k \in \{ 0,1, \dots, N_{DAC} \}
    \label{eq:llse_flin}
\end{equation}

\begin{equation}
    \mathbf{F} = \begin{bmatrix} \mathbf{F_{P}} & \mathbf{F_{1/P}} \end{bmatrix} \label{eq:llse_f}
\end{equation}

\begin{equation}
    \mathbf{F_{P}} = 
    \begin{bmatrix}
        1 & f[0] & f[0]^2 & \ldots & f[0]^P \\
        1 & f[1] & f[1]^2 & \ldots & f[1]^P \\
        \vdots & \vdots & \vdots & \ddots & \vdots \\
        1 & f[N] & f[N]^2 & \ldots & f[N]^P
    \end{bmatrix}
    \label{eq:llse_f_p}
\end{equation}

\begin{equation}
    \mathbf{F_{1/P}} = 
    \begin{bmatrix}
        f[0]^{1/2} & f[0]^{1/3} & \ldots & f[0]^{1/P} \\
        f[1]^{1/2} & f[1]^{1/3} & \ldots & f[1]^{1/P} \\
        \vdots & \vdots & \ddots & \vdots \\
        f[N]^{1/2} & f[N]^{1/3} & \ldots & f[N]^{1/P}
    \end{bmatrix}
    \label{eq:llse_f_1_p}
\end{equation}

\begin{equation}
    \mathbf{V_{PD}} = \mathbf{F} \mathbf{\hat{b}} \label{eq:llse_vpd}
\end{equation}

\begin{equation}
    \mathbf{V_{PD}} = \begin{bmatrix} V_{PD}[0] & V_{PD}[1] & \ldots & V_{PD}[N] \end{bmatrix}^T
    \label{eq:dac_vpd}
\end{equation}

Note that $\mathbf{F} \neq \mathbf{\hat{F}}$, as $\mathbf{F}$ is constructed from the ideal frequency values in \eqref{eq:llse_flin}, while $\mathbf{\hat{F}}$ is composed of the estimated VCO frequency values, $\hat{f}$. In \eqref{eq:llse_flin}, $B_{des}$ is the desired chirp bandwidth.  In \eqref{eq:llse_f_p} and \eqref{eq:llse_f_1_p}, $N = N_{DAC}$, which is the number of predistorted voltage increments actuated by the QDAC, operating at $f_{DAC}$:

\begin{equation}
    N_{DAC} = f_{DAC} \cdot T_{chirp} = T_{chirp} / T_{DAC}
    \label{eq:ndac}
\end{equation}

This process assumes that, to begin the chirp, the QDAC output is reset to the original $V[0]$ used in the ``chart" phase. In addition, periodic DAC updates are performed to provide uniform interpolation over the target voltage range. Indeed, \eqref{eq:llse_flin} is framed so as to perform an interpolative predistortion based on the charted data. By extending $B_{des}$ beyond the charted range, it is also possible to perform an extrapolative predistortion with potentially degraded accuracy.

\subsubsection{Linearization}

Converting \eqref{eq:dac_voltage} into a discrete form, $\mathbf{V_{PD}}$ uniquely prescribes the required IDAC control codes, $D_{DAC}[k]$:

\begin{equation}
    \begin{aligned}
    V_{tune}(kT) = V[0] + \frac{I_{LSB}}{C_{DAC}} \cdot T \cdot \sum_{i = 1}^{k} D_{DAC}[i] \\
    T = T_{DAC}, \; \forall k \in \{ 1,2, \dots, N_{DAC} \}
    \end{aligned}
    \label{eq:dac_voltage_discrete}
\end{equation}

Since the starting voltage is established by a switched reset rather than the IDAC, $V_{PD}[0]$ can be disregarded, while the values at other times can be calculated in matrix form, given that $V_{tune}(kT) = V_{PD}[k]$ is desired:

\begin{equation}
    \mathbf{V_{PD,1}} = \mathbf{V[0]} + K_{DAC} \cdot \mathbf{L_1} \mathbf{D_{DAC}}
    \label{eq:dac_update}
\end{equation}

\begin{equation}
    \mathbf{V_{PD,1}} = \begin{bmatrix} V_{PD}[1] & V_{PD}[2] & \ldots & V_{PD}[N] \end{bmatrix}^T
    \label{eq:dac_vpd1}
\end{equation}

\begin{equation}
    K_{DAC} = \frac{I_{LSB}}{C_{DAC}} \cdot T_{DAC} = \frac{I_{LSB}}{C_{DAC}} \cdot \frac{1}{f_{DAC}}
    \label{eq:dac_kdac}
\end{equation}

\begin{equation}
    \mathbf{L_1}
    = 
    \begin{bmatrix}
        1 & 0 & 0 & \ldots & 0 \\
        1 & 1 & 0 & \ldots & 0 \\
        1 & 1 & 1 & \ldots & 0 \\
        \vdots & \vdots & \vdots & \ddots & \vdots \\
        1 & 1 & 1 & \ldots & 1
    \end{bmatrix} 
    \label{eq:dac_L1_matrix}
\end{equation}

\begin{equation}
    \mathbf{D_{DAC}} = \begin{bmatrix} D_{DAC}[1] & D_{DAC}[2] & \ldots & D_{DAC}[N] \end{bmatrix}^T
    \label{eq:dac_ddac_def}
\end{equation}

Here, $N = N_{DAC}$, $\mathbf{V[0]}$ is an $N_{DAC} \times 1$ vector with all entries equal to $V[0]$, and $\mathbf{L_1}$ is an $N_{DAC} \times N_{DAC}$ lower triangular matrix of ones, which is full rank and therefore invertible. 

\begin{equation}
    \mathbf{D_{DAC}} = K_{DAC}^{-1} \cdot \mathbf{L_1}^{-1} (\mathbf{V_{PD,1}} - \mathbf{V[0]})
    \label{eq:dac_ddac_solve}
\end{equation}

Equations \eqref{eq:dac_update} and \eqref{eq:dac_ddac_solve} demonstrate a deterministic mapping between the generated predistorted voltage values and the IDAC control codes in the ``chirp" phase. The values in $\mathbf{D_{DAC}}$ must be rounded to be applied to the IDAC, introducing quantization error which is more severe for large $I_{LSB}$. 

Generally speaking, the backward model estimate, $\mathbf{\hat{b}}$, can be used to generate a linear chirp of any duration and bandwidth, provided that the VCO (through its tuning range) and the QDAC (through $I_{LSB}$, $C_{DAC}$, $f_{DAC}$, and its full-scale range in current and voltage) can support it. The relative simplicity of the invoked computation allows functional integration within existing digital signal processing (DSP) and control on-chip. This provides a powerful and adaptive tool for optimal chirp predistortion in systems targeting low chirp duration and wide bandwidth. This method can be generalized to other VCO modulation techniques, such as VDAC varactor tuning or digital cap bank control, by substituting the relevant code-to-voltage or code-to-frequency mapping (in place of the QDAC mapping presented from \eqref{eq:dac_voltage_discrete} onward, or as a wholesale replacement for the voltage-to-frequency mapping, respectively).

\begin{table}[!t]
\renewcommand{\arraystretch}{1.3}
\caption{Learned VCO Forward and Backward Coefficients \\ with $P = 5$, $N_{PSS} = 10$, and $N_{chart} = 100$}
\label{table_models}
\centering
\begin{tabular}{c|c||c|c}
\textbf{Forward} & \textbf{Value} & \textbf{Backward} &  \textbf{Value}\\
\hline
$a_0$ & 0.0001 & $b_0$ & 0.0000\\
\hline
$a_1$ & 0.3316 & $b_1$ & -3.4181\\
\hline
$a_2$ & 0.4631 & $b_2$ & 1.8010\\
\hline
$a_3$ & 1.7203 & $b_3$ & -1.2828\\
\hline
$a_4$ & -1.9810 & $b_4$ & 0.6372\\
\hline
$a_5$ & 0.5626 & $b_5$ & -0.1211\\
\hline
-- & -- & $b_{1/2}$ & 14.0534\\
\hline
-- & -- & $b_{1/3}$ & -29.5988\\
\hline
-- & -- & $b_{1/4}$ & 30.2043\\
\hline
-- & -- & $b_{1/5}$ & -11.3465\\
\end{tabular}
\end{table}

\section{Models and Analysis}

To evaluate the efficacy of the proposed system, predistortion of a realistic VCO tuning curve is simulated in MATLAB. The periodicity of the resulting chirp FM error is analyzed in the frequency domain, with clear implications identified for system ranging performance. A discrete time-domain FMCW ranging model is then constructed and simulated to analyze the effects of deterministic chirp nonlinearity and phase noise on the IF output spectrum. 

\subsection{Chirp Predistortion and Synthesis}

\begin{figure}[!t]
\centering
\includegraphics[width=3.5in]{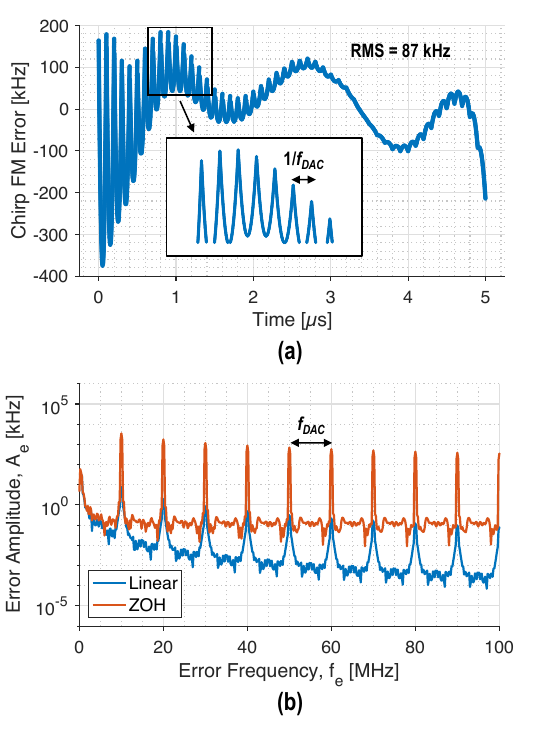}
\caption{(a) Chirp FM error in simulation from Fig. \ref{fig7} and (b) Hann-windowed 500000-point (whole-length) DFT of chirp FM error, comparing linear (blue) and ZOH (orange) DAC interpolation.}
\label{fig8}
\end{figure}

The complete learning, predistortion, and linearization process is illustrated in Fig. \ref{fig7}. The ``PSS Data" points in Fig. \ref{fig7}(a) come from Spectre Periodic Steady-State (PSS) simulations of a VCO implemented in a 28 nm CMOS process technology, including electromagnetic (EM) and post-layout extracted models. For these simulations, the tuning voltage of the VCO is swept in $N_{PSS} = 10$ steps of 0.1 V, resulting in $N_{PSS} + 1 = 11$ (voltage, frequency) pairs. The underlying VCO model for the frequency estimation and generation steps is derived from a fifth-order OLS polynomial fit of these data (OLS on \eqref{eq:llse_forward_model} with $P = 5$, Fig. \ref{fig7}(a)). The learned forward model coefficients are listed in the first two columns of Table \ref{table_models}, with no single term predominating. 

Learning is performed with $N_{chart} = 100$, assuming perfect frequency estimation. This isolates the DAC interpolation method, update frequency ($f_{DAC}$), and resolution as the sole contributors to deterministic chirp error at the output. The resulting learned coefficients for the backward model are listed in the last two columns of Table \ref{table_models}. The coefficients for the fractional polynomial terms are naturally larger, as they are reciprocal to the regular polynomial terms of the forward model. 

Chirp predistortion and synthesis are demonstrated in Figs. \ref{fig7}(b-d) for $T_{chirp} = $ 5 \textmu s, $f_{DAC} = $ 10 MHz, $I_{LSB} = $ 2.5 nA, and $C_{DAC} = $ 25 pF. A time step of 10 ps is used for all time-domain simulations in this work, with $I_{LSB}$ and $C_{DAC}$ unchanged throughout as well. These QDAC parameters reflect the capabilities of a current-steering architecture implemented in the same 28 nm CMOS process as the VCO. The use of a current mirror from the core IDAC to the output allows the effective resolution to scale down by the mirror ratio \cite{renukaswamy_12-mw_2020-2}. The chirp duration, $T_{chirp}$, can be reduced arbitrarily, with predistortion accuracy limited through $N_{DAC}$ by $f_{DAC}$ in \eqref{eq:ndac}. The chosen parameters result in 50 update steps, visualized in Fig. \ref{fig7}(b). Accuracy in the predistortion step (establishing $\mathbf{V_{PD}}$) can be improved by increasing $N_{DAC}$, typically through $f_{DAC}$. In a practical implementation, accuracy in the linearization step (realizing $V_{tune}(t)$) begins to degrade as $T_{DAC}$ approaches the settling time of the IDAC, due to dynamic settling error. Nonetheless, high $f_{DAC}$ has further benefits to chirp linearity, as will be detailed in the next section. 

\subsection{Chirp Nonlinearity Analysis}

\begin{figure*}[!t]
\centering
\includegraphics[width=7.16in]{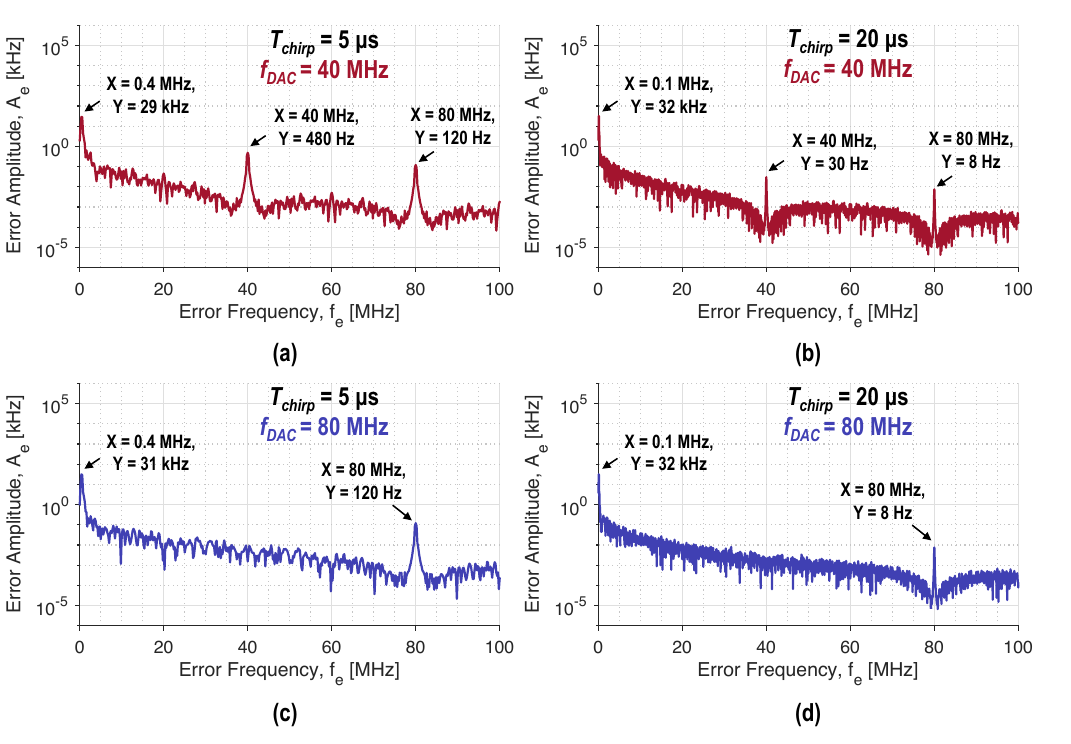}
\caption{Hann-windowed whole-length DFT of chirp FM error with linear DAC interpolation for (a) $T_{chirp} = $ 5 \textmu s, $f_{DAC} = $ 40 MHz; (b) $T_{chirp} = $ 20 \textmu s, $f_{DAC} = $ 40 MHz; (c) $T_{chirp} = $ 5 \textmu s, $f_{DAC} = $ 80 MHz; (d) $T_{chirp} = $ 20 \textmu s, $f_{DAC} = $ 80 MHz.}
\label{fig9}
\end{figure*}

In the absence of perfect DAC interpolation, periodic DAC updates to the tuning voltage naturally produce periodic components in the FM error of the resulting chirp. As depicted in Fig. \ref{fig8}, this error can be calculated with respect to a line of best fit, plotted, and analyzed. The Discrete Fourier Transform (DFT) of the real time-domain FM error signal, $e(t)$,  in Fig. \ref{fig8}(b) reveals that it is well-approximated with a sum of harmonics of $f_{DAC}$ in the amplitude-phase form:

\begin{equation}
    e(t) \approx e_{LF}(t) + \sum_{k = 1}^{\infty} e_{k}(t)
    \label{eq:chirp_error_decomp1}
\end{equation}

\begin{equation}
    e_{LF}(t) = A_{e,LF} \sin(2\pi f_{e,LF}t + \phi_{e,LF})
    \label{eq:chirp_error_LF}
\end{equation}

\begin{equation}
    e_{k}(t) = A_{e,k} \sin(2\pi f_{e,k}t + \phi_{e,k})
    \label{eq:chirp_error_decomp2}
\end{equation}

\begin{equation}
    f_{e,k} = k \cdot f_{DAC}
    \label{eq:chirp_error_decomp_fek}
\end{equation}

In \eqref{eq:chirp_error_decomp1} and \eqref{eq:chirp_error_decomp2}, $e_{k}(t)$ represents the error contribution from the $k$-th harmonic of $f_{DAC}$, and $A_{e,k}$  and $\phi_{e,k}$ represent the amplitude and phase of each component, respectively. A low-frequency error component, $e_{LF}(t)$, with associated parameters $A_{e,LF}$, $f_{e,LF} > 0$, and $\phi_{e,LF}$, emerges due to the shape of the VCO nonlinearity and is not mitigated by more accurate DAC interpolation.

Figs. \ref{fig8}(b) and \ref{fig9} show that the approximation in \eqref{eq:chirp_error_decomp1} is most valid for strong periodic FM error. This arises from less accurate DAC interpolation, either in the method (e.g., ZOH) or the number of points (low $N_{DAC}$). In addition, the error amplitude, $A_e$, is greatest for lower harmonics of $f_{DAC}$. Interestingly, when $f_{DAC}$ is scaled by a positive integer $Q$ to $Q \cdot f_{DAC}$, it is akin to removing the harmonics of $f_{DAC}$ which are not multiples of $Q \cdot f_{DAC}$, while harmonics of $Q \cdot f_{DAC}$ remain unchanged in magnitude. This has significant implications for the IF output spectrum and range detection, which are elaborated below using the framework of \cite{ayhan_impact_2016-1}.

First, the transmitted frequency ramp, $f_{TX}(t)$, can be expressed by adding $e(t)$ to the ideal chirp model of \eqref{eq:chirp_def}:

\begin{equation}
    f_{TX}(t) = f_{0} + \frac{B}{T_{chirp}} \cdot t + e(t)\label{eq:chirp_with_error}
\end{equation}

Random frequency variations arising from noise are analyzed separately as phase noise in subsequent sections. The output FMCW signal is then given in complex exponential form as

\begin{equation}
    s_{TX}(t) = A_{TX} \cdot \exp(j \cdot \psi(t))
    \label{eq:chirp_TX}
\end{equation}

where

\begin{equation}
    \psi(t) = 2\pi f_{TX}(t) \cdot t
    \label{eq:chirp_psi}
\end{equation}

and $A_{TX}$ is the output amplitude. For simplicity, the starting phase of the signal is assumed to be zero and $\tau_{target}$ is shortened to $\tau$. The IF output signal can then be calculated assuming ideal coherent mixing and low-pass filtering, with a TX-through-RX gain of $G_{RX}$ and net IF amplitude of $A_{IF}$:

\begin{equation}
    s_{IF}(t) = G_{RX} \cdot \text{LPF} \{ s_{TX}(t) \cdot s_{TX}(t - \tau) \}
    \label{eq:chirp_IF1}
\end{equation}

\begin{equation}
    s_{IF}(t) = A_{IF} \cdot \exp(j \cdot (\psi(t) - \psi(t - \tau))) 
    \label{eq:chirp_IF2}
\end{equation}

The argument in \eqref{eq:chirp_IF2} can be approximated for small $\tau$ (typical in radar) using the derivative of \eqref{eq:chirp_psi} \cite{ayhan_impact_2016-1}:

\begin{equation}
    s_{IF}(t) \approx A_{IF} \cdot \exp(j \cdot \frac{d\psi (t)}{dt} \cdot \tau) 
    \label{eq:chirp_IF3}
\end{equation}

\begin{equation}
    s_{IF}(t) \approx A_{IF} \cdot \exp(j \cdot 2 \pi f_{TX}(t) \cdot \tau)
    \label{eq:chirp_IF4}
\end{equation}

The argument can now be seen to consist of an expected component, $\psi_{des}(t)$, and an error component, $\psi_{e}(t)$:

\begin{equation}
    s_{IF}(t) \approx A_{IF} \cdot \exp(j \cdot (\psi_{des}(t) + \psi_{e}(t)))
    \label{eq:chirp_IF5}
\end{equation}

\begin{equation}
    \psi_{des}(t) = 2 \pi \tau \cdot (f_{0} + \frac{B}{T_{chirp}} \cdot t)
    \label{eq:chirp_psi_des}
\end{equation}

\begin{equation}
    \psi_{e}(t) = 2 \pi \tau \cdot e(t) \approx 2 \pi \tau \cdot ( e_{LF}(t) + \sum_{k = 1}^{\infty} e_{k}(t))
    \label{eq:chirp_psi_e}
\end{equation}

This allows the following factorization of $s_{IF}(t)$:

\begin{equation}
    s_{IF}(t) \approx A_{IF} \cdot s_{des}(t) \cdot s_{e,LF}(t) \cdot \prod_{k = 1}^{\infty} s_{e,k}(t)
    \label{eq:chirp_IF6}
\end{equation}

\begin{equation}
    s_{des / e,LF/e,k}(t) = \exp(j \cdot \psi_{des / e,LF/e,k}(t))
    \label{eq:chirp_sdes}
\end{equation}



\begin{equation}
    \psi_{e,LF}(t) = 2 \pi \tau \cdot e_{LF}(t), \; \psi_{e,k}(t) = 2 \pi \tau \cdot e_k(t)
    \label{eq:chirp_psi_ek}
\end{equation}

Each error factor can be expressed through the Jacobi-Anger series expansion using the Bessel functions of the first type, $J_n(z)$ \cite{wikipedia_jacobianger_2025, ayhan_impact_2016-1, bauduin_impact_2023}:

\begin{equation}
    \exp(j \cdot z \sin \theta) = \sum_{n = - \infty}^{\infty} J_n(z) \cdot \exp(j n \theta)
    \label{eq:jacobi_anger}
\end{equation}

\begin{equation}
    s_{e,LF}(t) = \sum_{m = - \infty}^{\infty} J_m(2 \pi \tau A_{e,LF}) \cdot \exp(j m (2 \pi f_{e,LF} + \phi_{e,LF})) 
    \label{eq:chirp_s_eLF_series}
\end{equation}

\begin{equation}
    s_{e,k}(t) = \sum_{n = - \infty}^{\infty} J_n(2 \pi \tau A_{e,k}) \cdot \exp(j n (2 \pi f_{e,k} + \phi_{e,k})) 
    \label{eq:chirp_s_ek_series}
\end{equation}

As a general result, these equations predict two phenomena that are commonly observed in the literature but typically treated only empirically: so-called spectral smearing and ghost targets \cite{samala_signal_2018}. The desired signal, $s_{des}(t)$, is henceforth assumed to be real with spectral content at $\pm f_{target}$. 

\subsubsection{Spectral Smearing}

\begin{figure}[!t]
\centering
\includegraphics[width=3.5in]{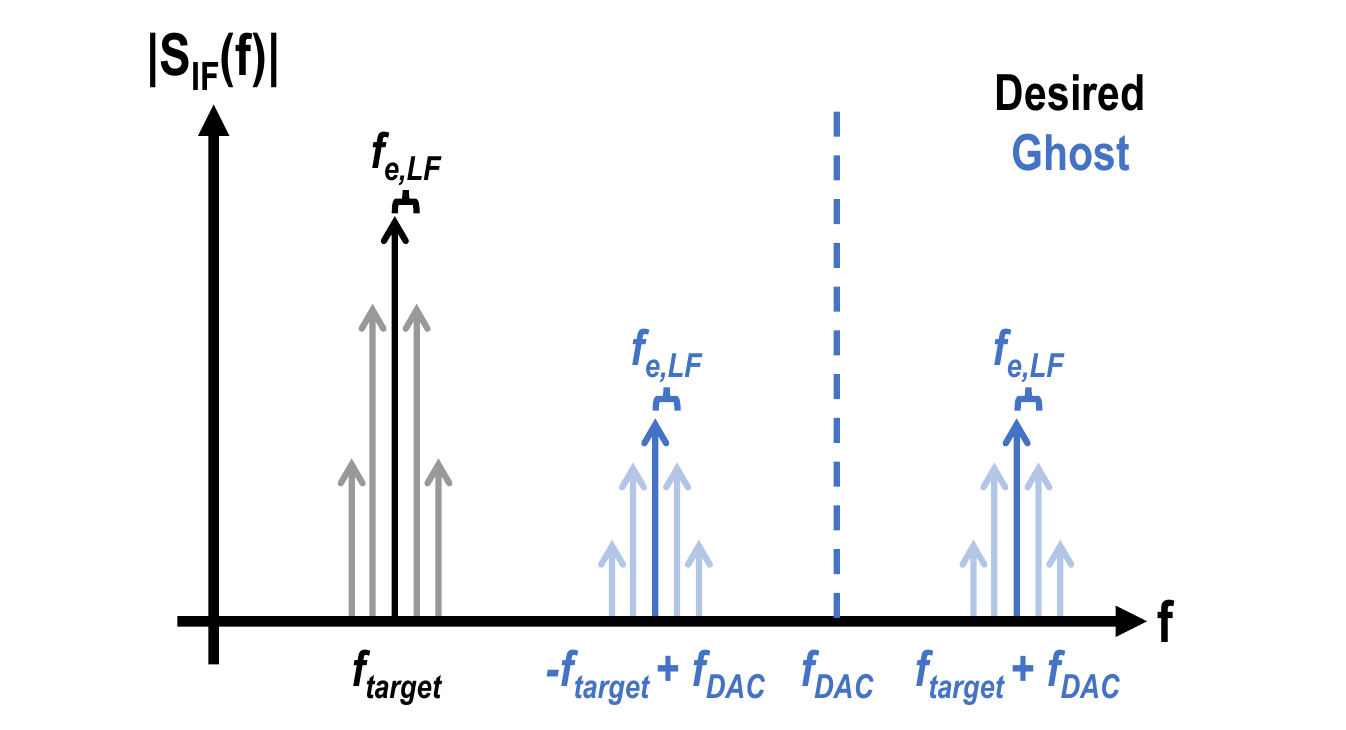}
\caption{Simplified IF spectrum with smearing (lightened) and ghost targets (blue) for $2f_{target} < f_{DAC}$.}
\label{fig10}
\end{figure}

First, the apparent width of the target lobe in the IF DFT increases when the chirp exhibits integral nonlinearity (INL), degrading the effective range resolution \cite{samala_signal_2018}. When the INL is viewed as a combination of slow-moving periodic error components in the chirp (e.g., $e_{LF}(t)$), it becomes clear based on the multiplied terms in \eqref{eq:chirp_IF6} and \eqref{eq:chirp_s_eLF_series} that the ``smearing" effect in the spectrum results analytically from the modulation products of $s_{e,LF}(t)$ with $s_{des}(t)$. Falling at $\pm f_{target} \pm m \cdot f_{e,LF} \; \forall m \in \mathbb{N}_0$ , these terms can be difficult to distinguish, since the DFT resolution is limited to the reciprocal of the observation time $T_{obs}$ (at best $1 / T_{overlap}$, Fig. \ref{fig1}), which is similar to $f_{e,LF}$ for short chirps. Adding the error contribution at other low frequencies around $f_{e,LF}$, which can be greater than that from the harmonics of $f_{DAC}$ (Fig. \ref{fig9}), the multitude of modulated terms close to the desired target introduces frequency ambiguity in the IF DFT.

\subsubsection{Ghost Targets}

Chirp differential nonlinearity (DNL) has been observed to cause additional peaks in the IF spectrum that do not correspond to physical targets, known as ``ghost targets" \cite{samala_signal_2018}. These spurious tones (spurs) arise deterministically from the higher frequency error components in the chirp, $e_k(t)$, manifesting in the IF output as $s_{e,k}(t)$. By \eqref{eq:chirp_error_decomp_fek} and \eqref{eq:chirp_s_ek_series}, $s_{e,k}(t) \; \forall k \in \mathbb{N}$ is composed exclusively of components at all integer multiples of $f_{DAC}$: $\pm n \cdot f_{DAC} \; \forall n \in \mathbb{N}_0$. The frequency coverage of the product of the $s_{e,k}(t)$ terms in \eqref{eq:chirp_IF6} is identical, as it is the result of replicating a spectrum only including components at $\pm n_1 \cdot f_{DAC} \; \forall n_1 \in \mathbb{N}_0$ about $\pm n_2 \cdot f_{DAC} \; \forall n_2 \in \mathbb{N}_0$ (repeatedly). The magnitude of $J_n(2 \pi \tau A_{e,k})$ decreases with $n$ for $2 \pi \tau A_{e,k} < 1$, causing the spectrum of this product to have a low-pass characteristic. 

When the product of $s_{e,k}(t)$ terms is multiplied by $s_{des}(t) \cdot s_{e,LF}(t)$ in \eqref{eq:chirp_IF6}, the smeared terms at $\pm f_{target} \pm m \cdot f_{e,LF} \; \forall m \in \mathbb{N}_0$ are scaled in amplitude and replicated in frequency about all integer multiples of $f_{DAC}$, resulting in terms at $\pm f_{target} \pm m \cdot f_{e,LF} \pm n \cdot f_{DAC} \; \forall m,n \in \mathbb{N}_0$ (Fig. \ref{fig10}). The low-pass nature of the product of $s_{e,k}(t)$ terms contributes to the ghost targets ``falling off" in magnitude about the desired peak in the spectrum. The smeared terms exhibit similar behavior about the desired and ghost targets. 

For $2 \pi \tau A_{e,k} < 1$, the following occurs as $A_{e,k}$ increases: the magnitude at the desired frequency, set by $J_0(2 \pi \tau A_{e,k})$, decreases, while the magnitude of the spurs, set by $J_n(2 \pi \tau A_{e,k}) \; \forall n > 0$, increases \cite{ayhan_impact_2016-1}. This point also applies to $A_{e,LF}$ and motivates accurate DAC interpolation to reduce the amplitude of all error components (Fig. \ref{fig8}(b)). 

Leaving aside the desired target tones and the smeared terms, the ghost targets can be isolated at $\pm f_{target} \pm n \cdot f_{DAC} \; \forall n \in \mathbb{N}$. Ghost targets can therefore be avoided within the desired IF band under the following condition:

\begin{equation}
    f_{max} < -f_{target} + f_{DAC} 
    \label{eq:IF_ghost1}
\end{equation}

\begin{equation}
    2f_{max} = 2 \cdot \frac{B}{T_{chirp}} \cdot \frac{2R_{max}}{c}  <  f_{DAC} 
    \label{eq:IF_ghost2}
\end{equation}

The maximum detectable range and its associated IF frequency are represented as $R_{max}$ and $f_{max}$, respectively. By satisfying the condition in \eqref{eq:IF_ghost2}, accurate range detection remains possible even for imprecise DAC interpolation, as long as spectral smearing effects (chirp INL) can be sufficiently mitigated at the macroscale by chirp predistortion \cite{liu_ultralow_2019-1}. All else being equal, this relation generally motivates making $f_{DAC}$ as high as possible during chirp generation. A practical limit arises at the point that the DAC interpolation accuracy degrades significantly due to $1/f_{DAC}$ approaching or dropping below the settling time of the DAC.

\begin{figure}[!t]
\centering
\includegraphics[width=3.5in]{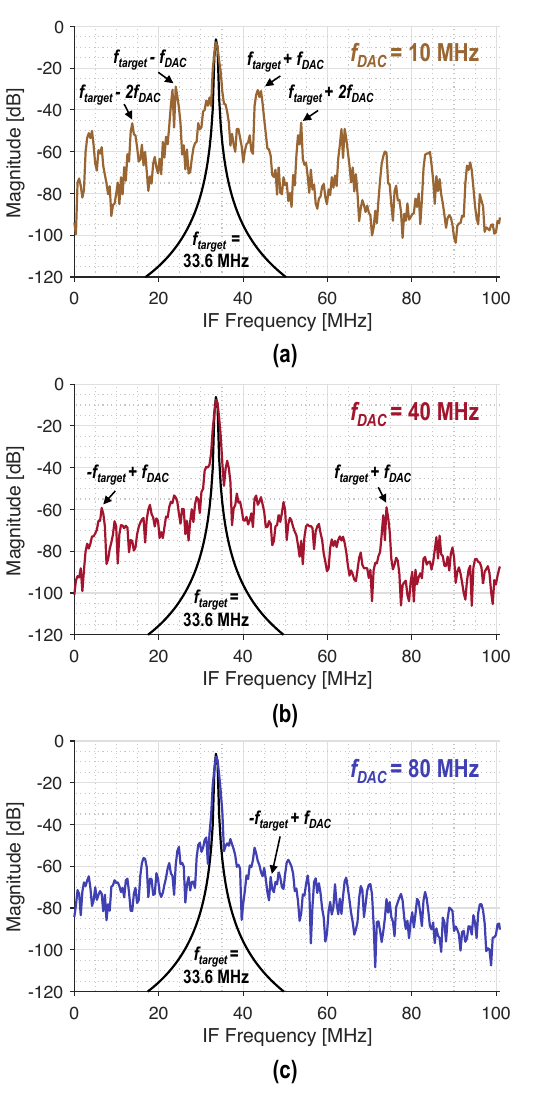}
\caption{Single target Hann-windowed 262144-point IF DFT for $R = $ 23 m and $T_{chirp} = $ 5 \textmu s, for (a) $f_{DAC} = $ 10 MHz, (b) $f_{DAC} = $ 40 MHz, and (c) $f_{DAC} = $ 80 MHz, with linear DAC interpolation and no phase noise added. The result for an ideal chirp is shown in black. Each DFT starts 40\% into $T_{overlap}$.}
\label{fig11}
\end{figure}

\begin{figure}[!t]
\centering
\includegraphics[width=3.5in]{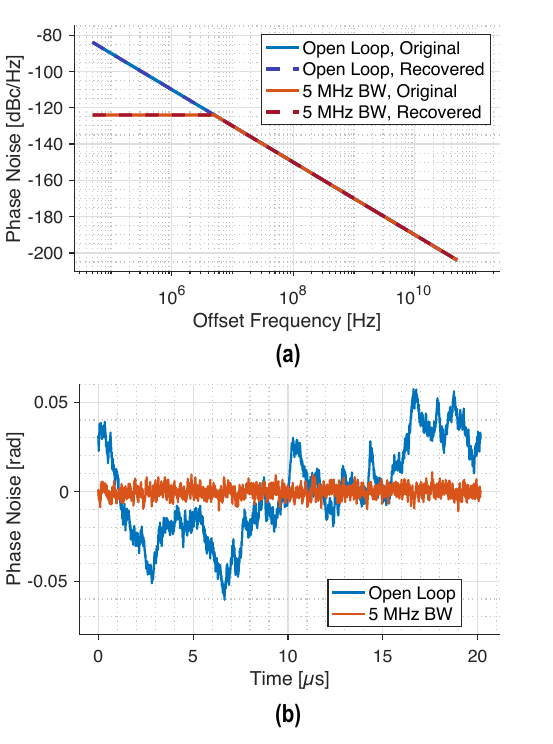}
\caption{TX phase noise (a) original (solid) and recovered (dashed) spectra, with (b) example generated time series $\phi_n[k]$, for $T_{chirp} = $ 20 \textmu s.}
\label{fig13}
\end{figure}

\begin{figure}[!t]
\centering
\includegraphics[width=3.5in]{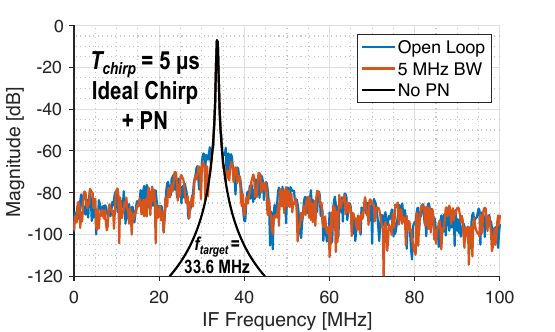}
\caption{Single target Hann-windowed 480000-point IF DFT for $R = $ 23 m and $T_{chirp} = $ 5 \textmu s, with varying levels of phase noise added to an ideal chirp (Fig. \ref{fig13}). Each DFT starts from the beginning of $T_{chirp}$.}
\label{fig12}
\end{figure}

\begin{figure}[!t]
\centering
\includegraphics[width=3.5in]{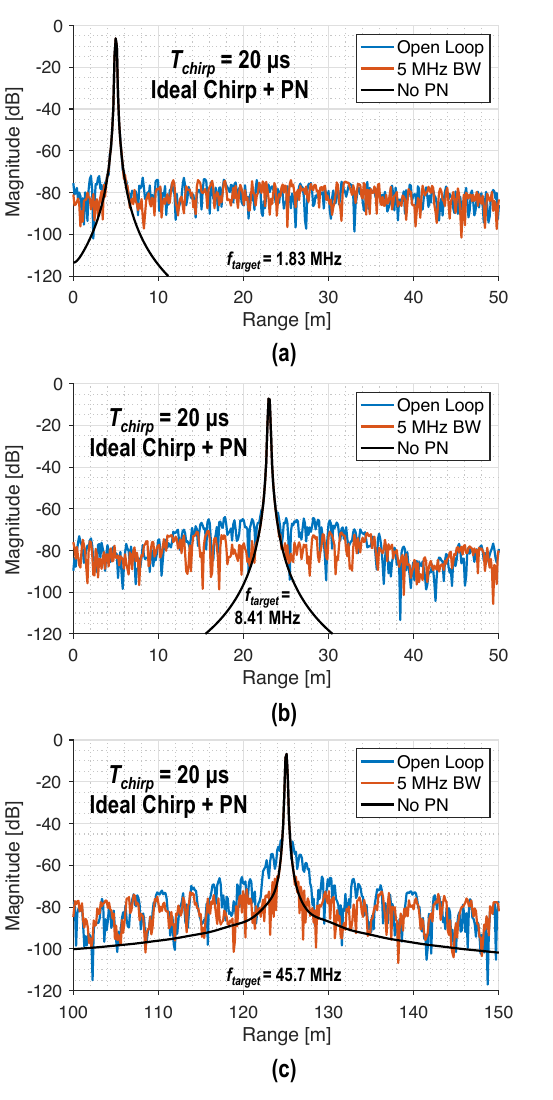}
\caption{Single target Hann-windowed 1980000-point IF DFT for $R = $ (a) 5 m, (b) 23 m, and (c) 125 m and $T_{chirp} = $ 20 \textmu s, with varying levels of phase noise added to an ideal chirp (Fig. \ref{fig13}). Each DFT starts from the beginning of $T_{overlap}$.}
\label{fig14}
\end{figure}

\begin{figure}[!t]
\centering
\includegraphics[width=3.5in]{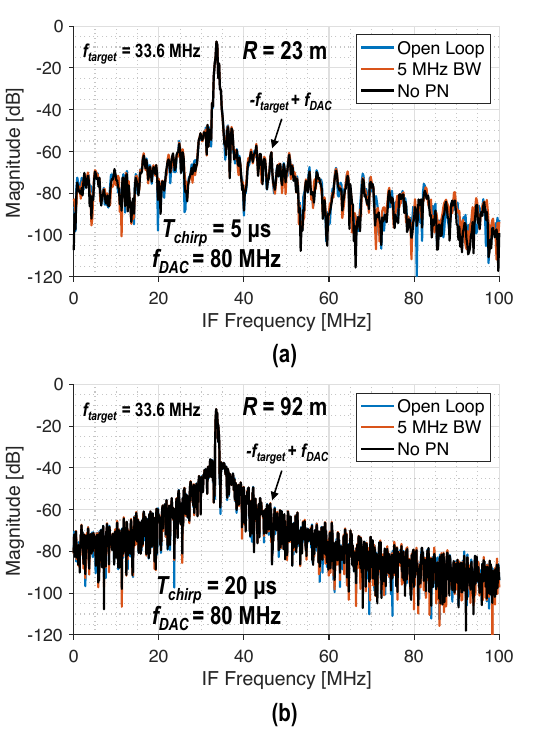}
\caption{ (a) Single target Hann-windowed 480000-point IF DFT for $f_{DAC} = $ 80 MHz, with $R = $ 23 m and $T_{chirp} = $ 5 \textmu s, above (b) similar 1980000-point IF DFT, with $R = $ 92 m and $T_{chirp} = $ 20 \textmu s. Each chirp is generated by linear DAC interpolation, with varying levels of phase noise added (Fig. \ref{fig13}). Each DFT starts from the beginning of $T_{overlap}$.}
\label{fig15}
\end{figure}

\begin{figure}[!t]
\centering
\includegraphics[width=3.5in]{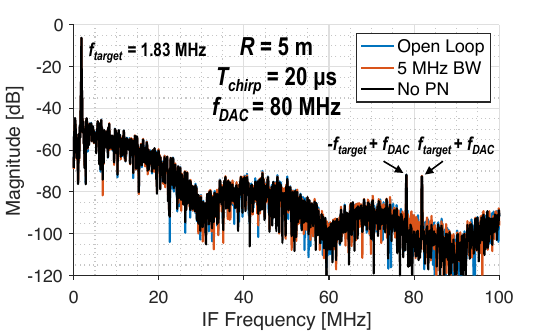}
\caption{Single target Hann-windowed 1980000-point IF DFT for $R = $ 5 m, $T_{chirp} = $ 20 \textmu s, and $f_{DAC} = $ 80 MHz, with linear DAC interpolation and varying levels of phase noise added (Fig. \ref{fig13}). Each DFT starts from the beginning $T_{overlap}$.}
\label{fig16}
\end{figure}

\subsection{Coherent Mixing and Phase Noise}

As expressed in \eqref{eq:chirp_IF1}, the monostatic FMCW radar system features coherent mixing, in which the two input signals to the mixer have a fixed phase offset \cite{siddiq_phase_2019-1}. Coherent mixing produces the following relation between the TX and IF phase noise ($\mathcal{L}_{TX}$ and $\mathcal{L}_{IF}$), known as the range correlation effect \cite{budge_range_1993, stove_linear_1992-2, siddiq_phase_2019-1}:

\begin{equation}
    \mathcal{L}_{IF}(\Delta f) = \mathcal{L}_{TX}(\Delta f) \cdot 4 \sin^2(\pi \tau \cdot \Delta f)
    \label{eq:IFPN}
\end{equation}

The IF phase noise is the TX phase noise multiplied by a factor, $4 \sin^2(\pi \tau \cdot \Delta f)$, which is dependent on the time-of-flight delay, $\tau$, and the offset frequency, $\Delta f$. Phase noise cancellation at the IF relative to the TX is observed when this multiplicative factor is less than one, equivalent to $\tau \cdot \Delta f < 1/6$ \cite{siddiq_phase_2019-1}. The envelope exhibits extrema at $\Delta f = k/(2 \tau) \; \forall k \in \mathbb{N}$, with nulls at even $k$ and peaks at odd $k$. The IF phase noise rises 6 dB (a factor of 4) in the peak case. This effect can be seen clearly in the IF spectra presented in Figs. \ref{fig12} and \ref{fig14}, which are simulated for a single target with ideal, perfectly linear chirps. 

The combined effects of chirp nonlinearity and phase noise on the IF spectrum are modeled in MATLAB using the following real, discrete-time equivalent equations:

\begin{equation}
    s_{TX}[k] = \cos(2 \pi \cdot f_{chirp}[k] \cdot t[k] + \phi_n[k])
    \label{eq:model_sTX}
\end{equation}

\begin{equation}
    s_{RX}[k] = \cos(2 \pi \cdot f_{chirp}[k - q_{\tau} ] \cdot t[k] + \phi_n[k - q_{\tau} ])
    \label{eq:model_sRX}
\end{equation}

\begin{equation}
    s_{IF}[k] = 2 \cdot s_{TX}[k] \cdot s_{RX}[k]
    \label{eq:model_sIF}
\end{equation}

The output is valid for non-negative integers $k \geq q_{\tau}$, where $q_{\tau} = \lfloor \tau / T \rfloor$, $\tau$ is again the time-of-flight delay, and $T$ is the time step. The chirp frequency values, $f_{chirp}[k]$ are generated through the ``chart \& chirp" procedure described previously. The phase noise values, $\phi_n[k]$, are produced through the following process in MATLAB, with vectors presented alongside their index variables and arithmetic operations executed pointwise \cite{hilmar_answer_2018, noauthor_ifft_nodate, rubiola_companion_2023-1}:

\begin{enumerate}

\item{$N$ is the desired number of points in the time domain. $N$ is chosen odd such that $L = (N-1)/2$ is an integer.}

\item{The desired, real-valued single sideband (SSB) phase noise $\mathcal{L}_{TX}[\Delta f]$ is defined over $L$ points for $\Delta f_{low} \leq \Delta f \leq \Delta f_{high}$, with units of dBc/Hz.}

\item{$S_{\phi}[\Delta f] / 2 = 10^{\mathcal{L}_{TX}[\Delta f]/10}$, with units of $\text{rad}^2/\text{Hz}$}

\item{The resolution bandwidth $\text{RBW} = 2\Delta f_{high}/N$.}

\item{The double sideband (DSB) magnitude spectrum, with units of rad, is constructed at $[0, + \Delta f, - \Delta f]$ as $[0, \sqrt{\text{RBW} \cdot S_{\phi}[\Delta f] / 2}, \sqrt{\text{RBW} \cdot \text{flip} (S_{\phi}[\Delta f] / 2)}]$. }

\item{A unity-gain SSB random phase vector is created as $K_{\phi}[\Delta f] = \exp(j \cdot 2 \pi \cdot \text{rand}(1,L))$.}

\item{This SSB vector is converted into conjugate-symmetric DSB form as $[1, K_{\phi}[\Delta f], \text{conj(flip(} K_{\phi}[\Delta f]))]$. }

\item{The DSB magnitude spectrum is multiplied pointwise by the DSB random phase vector and then scaled by $N$ to produce $S_{\phi,DSB}[0, + \Delta f, - \Delta f]$.}

\item{$\phi_n[k] = \text{real(ifft(} S_{\phi,DSB}[0, + \Delta f, - \Delta f]))$.}

\end{enumerate}

TX phase noise spectra $\mathcal{L}_{TX}[\Delta f]$ and time series $\phi_n[k]$ are constructed as shown in Fig. \ref{fig13} and described below. The spectra of the time series $\phi_n[k]$ are recovered via one-sided periodogram (with rectangular windowing) and confirmed to match the original constructed spectra \cite{rubiola_companion_2023-1}. Spectre PNOISE simulation results for the designed VCO (Fig. \ref{fig7}) establish a worst case TX phase noise of -110 dBc/Hz at 1 MHz offset with respect to the maximum VCO frequency, 9.741 GHz. Assuming negligible $1 / (\Delta f)^3$ phase noise, an ``Open Loop" phase noise spectrum with a $1 / (\Delta f)^2$ shape is created to cross the simulated worst case level. A ``5 MHz BW" spectrum, which matches the ``Open Loop" spectrum beyond 5 MHz but maintains a phase noise level of -124 dBc/Hz for lower offsets, is created to emulate the ``phase noise pedestal" effect of many closed-loop (i.e., PLL) frequency synthesizers \cite{siddiq_phase_2019-1}. The bounds of the SSB spectrum are constrained at the low end by the observation time $T_{obs} \leq T_{chirp}$ and at the high end by the simulation time step, $\Delta t$: $\Delta f_{low} = 1 / T_{obs}$ and $\Delta f_{high} = 1 / (2 \Delta t)$. A practical benefit of the lower bound is that less low-frequency phase noise is manifested for short chirps.

\section{Results and Discussion}

\begin{figure}[!t]
\centering
\includegraphics[width=3.5in]{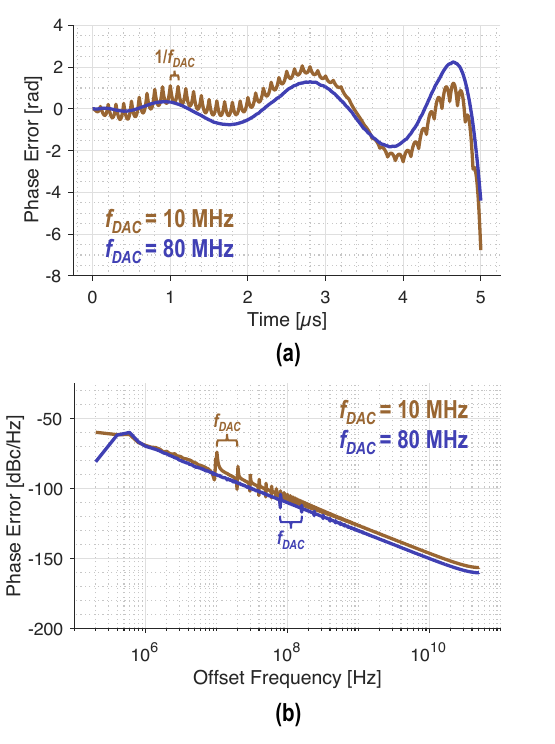}
\caption{Phase error (a) time series and (b) recovered spectra for $T_{chirp} = $ 5 \textmu s and $f_{DAC} = $ 10 MHz and 80 MHz, with linear DAC interpolation.}
\label{fig17}
\end{figure}

\begin{figure}[!t]
\centering
\includegraphics[width=3.5in]{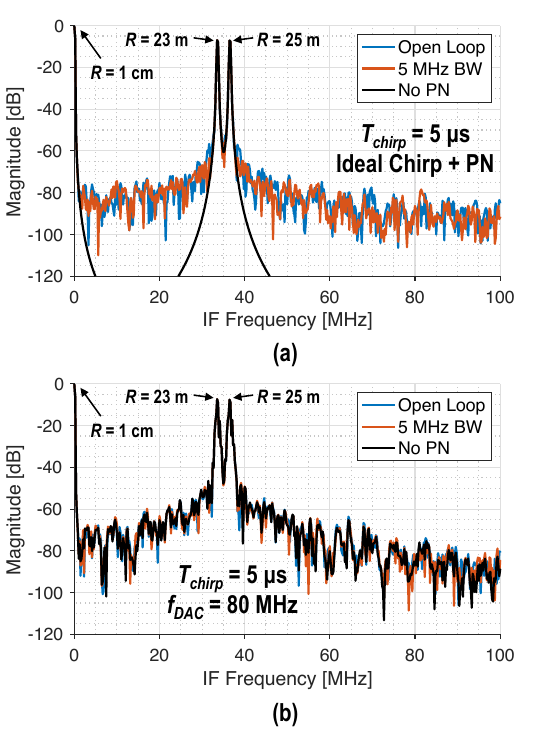}
\caption{Multiple target Hann-windowed 480000-point IF DFT for $R = $ 1 cm, 23 m, and 25 m, with $T_{chirp} = $ 5 \textmu s and varying levels of phase noise added (Fig. \ref{fig13}). Each chirp in (a) is ideal, while each chirp in (b) is generated through linear DAC interpolation with $f_{DAC} = $ 80 MHz. Each DFT starts from the beginning of $T_{overlap}$.}
\label{fig18}
\end{figure}

The effect of chirp nonlinearity on the IF spectrum in the absence of random phase noise is visualized for the single target case in Fig. \ref{fig11}, where the IF spectra resulting from the ``chart \& chirp" process with varying $f_{DAC}$ are compared to the IF spectrum resulting from an ideal chirp. In all cases, the main lobe at the target frequency is apparently widened and smeared, and ghost targets appear at the predicted frequencies. Verifying the prior analysis, the spurs resulting from successively higher harmonics of $f_{DAC}$ ``fall off" in magnitude (Fig. \ref{fig11}(a)). Increasing $f_{DAC}$ improves the DAC interpolation accuracy and decreases $A_{e,k}$ for the harmonic error terms, lowering the magnitude of the ghost targets. Increased nonlinearity from the early portion of the VCO tuning curve (e.g., Fig. \ref{fig8}(a)) is avoided here by starting the DFT 40\% into the TX and RX chirp overlap window, $T_{overlap}$ (Fig. \ref{fig1}). This is similar to initializing the VCO tuning voltage at a nonzero value, as is done in practice \cite{renukaswamy_12-mw_2020-2}. This approach reduces the magnitude of the ghost targets and increases the magnitude of the real target, improving the SNDR when the spurs are strong, at the expense of a higher DFT noise floor due to lower $T_{obs}$ and increased DFT bin sizing.  

Although the many low-frequency error terms evident in Figs. \ref{fig8}(b) and \ref{fig9} are reduced to just one, at $f_{e,LF}$, for simplicity in the preceding analysis, it is clear from Fig. \ref{fig11} that the interaction of these and other neglected terms contributes to a smearing ``skirt" that extends far beyond the target --- contrary to the simplified depiction in Fig. \ref{fig10}. In fact, a comparison of Figs. \ref{fig11} and \ref{fig12} indicates that this nonlinearity-induced skirt may exceed any skirt induced by phase noise for the VCO represented in this work. 

The results for $T_{chirp} =$ 5 \textmu s in Figs. \ref{fig11} and \ref{fig15}(a) reveal that the nonlinearity-induced skirt can have nulls which match those seen in the skirts induced by phase noise in Fig. \ref{fig12}. This points to a situation where non-periodic FM error behaves more like phase noise and is therefore also shaped by the range correlation effect. Indeed, Fig. \ref{fig16} confirms this, as the expected range correlation nulls based on \eqref{eq:IFPN} at 31.8 MHz, 61.8 MHz, and 91.8 MHz are clearly apparent. This behavior can be dissected analytically by expressing the FM error in the TX chirp, $e(t)$, in terms of phase error, $\phi_e(t)$:

\begin{equation}
    s_{TX}(t) = A_{TX} \cdot \exp(j \cdot ( 2 \pi \cdot f_{chirp}(t) \cdot t + \phi_e(t)))
    \label{eq:chirp_TX_phie}
\end{equation}

\begin{equation}
    \phi_e(t) = 2 \pi \cdot e(t) \cdot t
    \label{eq:chirp_phie}
\end{equation}

The phase error time series and spectra recovered via one-sided periodogram (with rectangular windowing) for $f_{DAC} = $ 10 MHz and 80 MHz are plotted in Fig. \ref{fig17}. Both spectra exhibit a -20 dB/decade slope, resembling $1 / (\Delta f)^2$ phase noise but at a significantly higher level than that from the VCO. The approximately 40 dB difference between this deterministic phase error (-70 dBc/Hz at 1 MHz offset) and the random phase noise (-110 dBc/Hz at 1 MHz offset for ``Open Loop") explains why the simulated IF spectra are largely unperturbed by the inclusion of random phase noise alongside chirp nonlinearity in the model. 

The recovered phase error spectra also feature spurs at the harmonics of $f_{DAC}$, and these are much stronger for $f_{DAC} = $ 10 MHz. This verifies the previous analysis of periodic FM error, as well as the results in Fig. \ref{fig11}(a). As a tool, the recovered spectrum of the chirp phase error can be used to identify whether the noise and distortion floor of the IF DFT will be dominated by non-periodic FM error or random phase noise, as well as to predict the location and magnitude of spurs arising due to periodic FM error.  

Nonetheless, the IF spectra resulting from the ``chart \& chirp" process with sufficiently high $f_{DAC}$ have high SNDR. An SNDR of about 24 dB is achieved in the single target case of Fig. \ref{fig15}(b), which is representative of target detection at long range such that $\tau \cdot \Delta f > 1/2$ for close offsets. At closer range (Figs. \ref{fig15}(a) and \ref{fig16}), around 40 dB of SNDR is demonstrated, limited primarily at close offsets by deterministic phase error. 


A multiple target scenario ($R_1 =$ 23 m and $R_2 = $ 25 m) including TX-to-RX self-interference ($R_{SI} = $ 1 cm) is presented in Fig. \ref{fig18}. The range correlation effect ensures that the self-interference cannot obscure nearby targets. Self-interference remains a design concern, as it can desensitize the receiver \cite{chen_140ghz_2025}, but it does not otherwise adversely affect target detection, since the DFT range resolution $(1/T_{obs}) \cdot (c/2) \cdot (T_{chirp}/B) = $ 14.25 cm $>$ 1 cm $= R_{SI}$. 

The interaction of the nearby targets, modeled to produce equal amplitude at the receiver, illustrates the potential for degraded target distinction due to target smearing. The targets overlap in Fig. \ref{fig18}(b) at a level more than 10 dB higher than that of the ideal case in Fig. \ref{fig18}(a). In practice, a target detection algorithm, such as the constant false alarm rate (CFAR) algorithm, can call for more than 15 dB of SNDR \cite{wang_w-band_2025, bauduin_impact_2023}. 

It is typical to average the IF DFT results across multiple chirps to lower any DFT floor arising from random noise. However, deterministic phase error from the chirp nonlinearity is dominant and cannot be addressed through averaging, as it will repeat in each chirp. This issue is likely shared by other approaches to chirp synthesis reported in the literature, although this is difficult to confirm, since simulations or measurements of the IF spectrum during target detection are often omitted. Usually, the RMS of the chirp FM error is the only chirp nonlinearity metric that is reported. As presented in Table \ref{table_rms}, the ideal ``chart \& chirp" process compares favorably in this aspect against the measured state of the art. 

\begin{table}[!t]
\renewcommand{\arraystretch}{1.3}
\caption{Sawtooth Short Chirp RMS FM Error Comparison}
\label{table_rms}
\centering
\begin{tabular}{c|c||c|c|c}
\textbf{} & \textbf{This Work} & \textbf{\cite{renukaswamy_12-mw_2020-2}} &  \textbf{\cite{tesolin_10-ghz_2024-2}} & \textbf{\cite{zhang_276_2026}}\\
\hline
$f_{DAC}$ [MHz] & 80 & 80 & 250 & 100 \\
\hline
$B$ [GHz] & 1.097 & 1.21 & 0.647 & 0.89\\
\hline
$T_{chirp}$ [\textmu s] & 5, 20 & 12.8, 25.6 & 1 & 0.57\\
\hline
RMS [kHz] & 50, 51 & 120, 55 & 151 & 851\\
\hline
$N_{chart}$ & 100 & N/A, LMS & N/A, LMS & N/A, LMS\\

\end{tabular}
\end{table}

\section{Conclusion}

To conclude, this article presents the architecture and concepts behind the ``chart \& chirp" method of one-shot, in situ VCO tuning curve estimation, learning, and predistortion. First, the operation of the core blocks, including the cycle-counting FDC and linearly interpolating QDAC, is described and analyzed. Then, the LLSE-based learning process translating the estimated VCO tuning curve into predistorted DAC control codes is detailed. Through analysis of the DFT of the FM error of generated chirps, the locations of nonlinearity-induced spurs in the IF spectrum are predicted. In particular, it is identified that increasing the frequency of chirp update steps, $f_{DAC}$, results in higher interpolation accuracy, improved linearity, and weaker, higher frequency ghost targets at the IF.

A coherent, monostatic system model, including the effects of chirp nonlinearity and phase noise, is then constructed to simulate range sensing with the generated chirps in the time domain, showing about 40 dB and 24 dB of SNDR for $R \leq$ 23 m and $R =$ 92 m, respectively. Simulated results demonstrate that deterministic phase error arising from the chirp nonlinearity dominates random phase noise by up to 40 dB, setting the SNDR for target detection --- a general result likely applicable to all chirp generation methods. The ``chart \& chirp" method successfully mitigates periodic FM error components to achieve state-of-the-art linearity for fast chirps; however, the non-periodic origin of the problematic phase error indicates that analog nonlinearity compensation could potentially be used to further linearize the VCO tuning curve prior to estimation, learning, and predistortion. 


%





\ifCLASSOPTIONcaptionsoff
  \newpage
\fi




\bibliographystyle{IEEEtran}

\bibliography{bibtex/bib/IEEEabrv, bibtex/bib/references}
%



%

\begin{IEEEbiography}[{\includegraphics[width=1in,height=1.25in,clip,keepaspectratio]{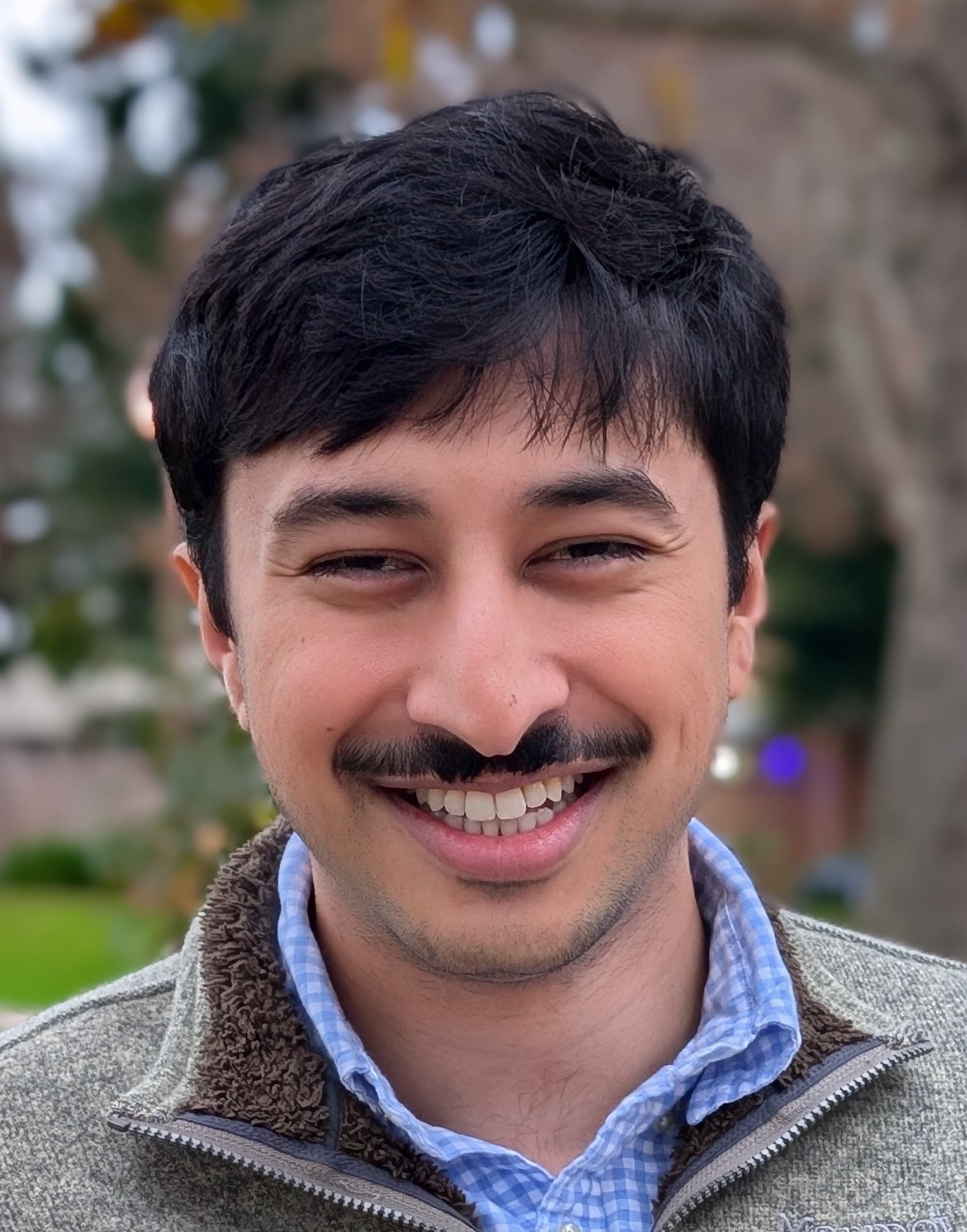}}]{Averal N. Kandala}
 received the B.S. (highest honors) and M.S. degrees in electrical engineering and computer sciences from the University of California (UC), Berkeley, in 2020 and 2021, respectively. He is currently a candidate for the Ph.D. degree at the same institution. He received a UC Berkeley Outstanding Graduate Student Instructor Award in 2025, as well as an NSF Graduate Research Fellowship in 2020. His research interests include integrated circuits and systems for sensing and communication.
\end{IEEEbiography}

\begin{IEEEbiography}[{\includegraphics[width=1in,height=1.25in,clip,keepaspectratio]{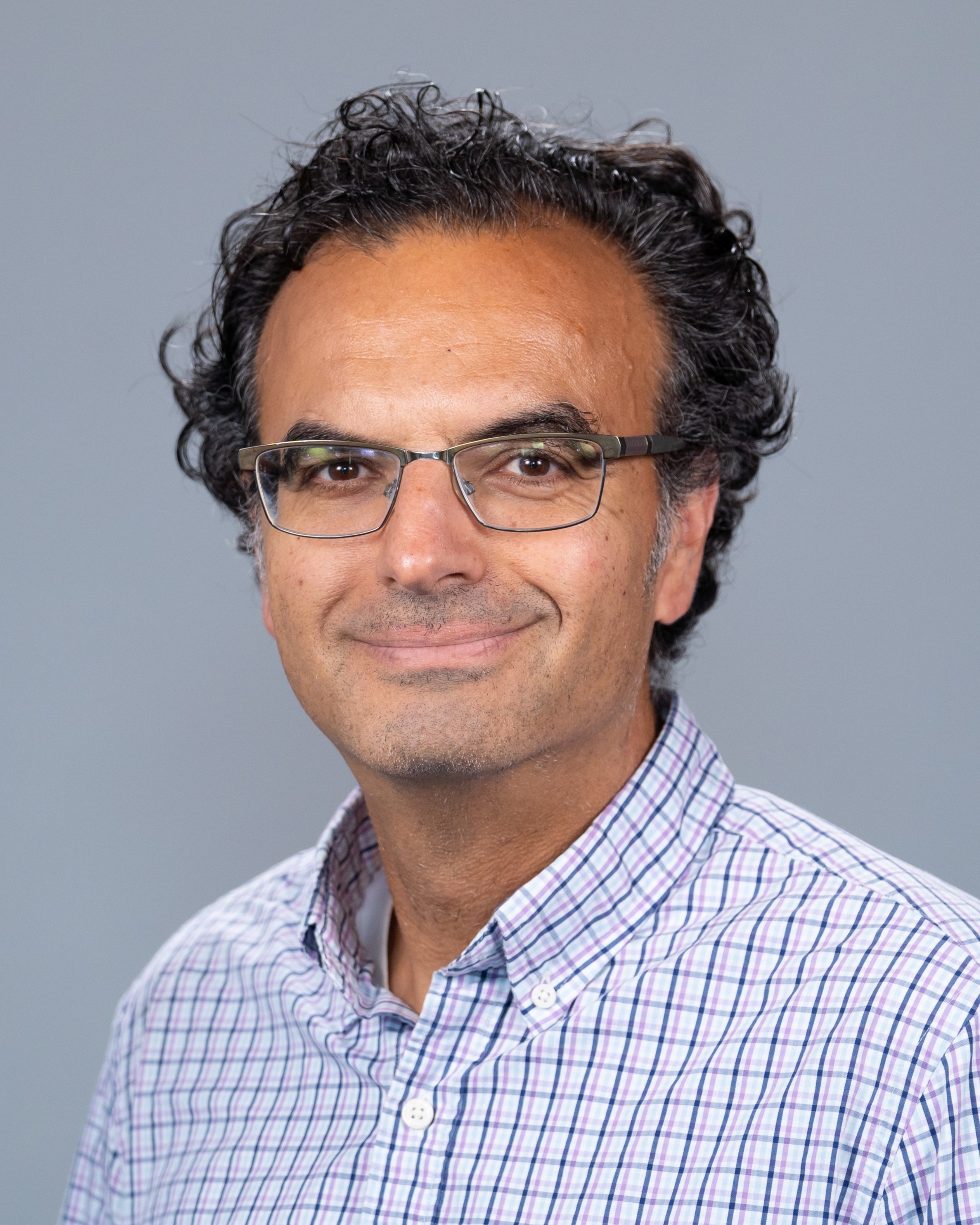}}]{Ali M. Niknejad}
 received the Ph.D. degree in electrical engineering from the University of California (UC), Berkeley, in 2000. He currently holds the Donald O. Pederson Distinguished Professorship chair in the EECS department at UC Berkeley, and he is a faculty co-director of the Berkeley Wireless Research Center (BWRC). He is also the Associate Director of the Center for Ubiquitous Connectivity (CUbiC) and served as the Associate Director for the Center for Converged TeraHertz Communications and Sensing (ComSenTer). He received the 2020 SIA/SRC University Research Award, recognized “for noteworthy achievements that have advanced analog, RF, and mm-wave circuit design and modeling, which serve as the foundation of 5G+ technologies.” 
\end{IEEEbiography}





\end{document}